\documentclass{svjour3}                     
\smartqed  
\usepackage{graphicx}
\begin{document}

\title{Point-Form Hamiltonian Dynamics and Applications%
\thanks{\lq\lq Relativistic Description of Two- and Three-Body Systems
in Nuclear Physics",\\ ECT*, October 19-13, 2009} }


\author{E.P. Biernat         \and
        W.H. Klink \and
        W. Schweiger 
}


\institute{E.P. Biernat \and W. Schweiger \at
              Institut f\"ur Physik, FB Theoretische Physik, Universit\"at Graz, 8010 Graz,
              Austria\\
              \email{elmar.biernat@uni-graz.at} \\
              \email{wolfgang.schweiger@uni-graz.at}
           \and
           W.H. Klink \at
           Department of Physics and Astronomy, University Iowa,
           Iowa City, IA, USA\\
           \email{william-klink@uiowa.edu}
}

\date{Received: date / Accepted: date}

\maketitle

\begin{abstract} This short review summarizes recent developments
and results in connection with point-form dynamics of relativistic
quantum systems. We discuss a Poincar\'e invariant multichannel
formalism which describes particle production and annihilation via
vertex interactions that are derived from field theoretical
interaction densities. We sketch how this rather general formalism
can be used to derive electromagnetic form factors of confined
quark-antiquark systems. As a further application it is explained
how the chiral constituent quark model leads to hadronic states
that can be considered as bare hadrons dressed by meson loops.
Within this approach hadron resonances acquire a finite
(non-perturbative) decay width. We will also discuss the
point-form dynamics of quantum fields. After recalling basic facts
of the free-field case we will address some quantum field
theoretical problems for which canonical quantization on a
space-time hyperboloid could be advantageous.

\end{abstract}

\section{Introduction}
\label{intro}
The formal problem of constructing a relativistically invariant
quantum theory amounts to finding a representation of the
Poincar\'e generators in terms of self-adjoint operators that act
on an appropriate Hilbert space and satisfy the Poincar\'e
algebra. If one is interested in interacting theories this is a
non-trivial task. The commutation relation
\begin{equation}
[\hat{P}^0, \hat{K}^j]=i \hat{P}^j
\end{equation}
between the generator for time translations $\hat{P}^0$ and the
boost generators $\hat{K}^j$ already indicates that interaction
terms must appear in more than one of the Poincar\'e generators.
If interaction terms show up in $\hat{P}^0$, then $\hat{K}^j$ or
$\hat{P}^j$ (or both) must be interaction dependent too. A
particular form of relativistic dynamics is then characterized by
the set of Poincar\'e generators that contain interactions. In
Dirac's seminal paper on the forms of classical relativistic
dynamics the interaction dependent (dynamical) Poincar\'e
generators were called \lq\lq Hamiltonians"~\cite{Dirac:1949cp}.
The most prominent forms are the \lq\lq instant form" (IF), the
\lq\lq front form" (FF), and the \lq\lq point form" (PF) with the
corresponding sets of Hamiltonians being given by $\{\hat{P}^0,
\hat{K}^1, \hat{K}^2, \hat{K}^3 \}$, $\{\hat{P}^{-} =
\hat{P}^0-\hat{P^3}, \hat{F}^1 = \hat{K}^1 - \hat{J}^2, \hat{F}^2
= \hat{K}^2 + \hat{J}^1\}$, and $\{\hat{P}^0, \hat{P}^1,
\hat{P}^2, \hat{P}^3 \}$, respectively. Among these forms the
point form is the least known and, although it has definite
virtues, the least utilized.
 An obvious advantage is the clean separation
of Poincar\'e generators that are interaction dependent from those
which are free of interactions. The former are components of a
4-vector $\hat{P}^\mu$ whereas the latter can be combined into an
antisymmetric Lorentz tensor $\hat{J}^{\mu\nu}$. This makes it
possible to express equations in point form in a manifestly
Lorentz covariant way and makes for simple behavior under
rotations and boosts. The conditions for Poincar\'e invariance
can, e.g., be phrased in terms of the \lq\lq point-form equations"
\begin{eqnarray}
[\hat{P}^\mu,\hat{P}^\nu]&=&0\, , \nonumber\\
\hat{U}_\Lambda \hat{P}^\mu \hat{U}_\Lambda^\dag &=& \left(
\Lambda^{-1} \right)^\mu_\nu \hat{P}^\nu \, ,
\label{eq:pfeq}\end{eqnarray} where $\hat{U}_\Lambda$ is the
unitary operator representing the Lorentz transformation $\Lambda$
on the Hilbert space.  The Lorentz part of the Poincar\'e
commutation relations has been integrated out, which is possible
due to the kinematic nature of the rotation and boost generators.

The usual procedure for constructing a representation of the
Poincar\'e algebra on a Hilbert space is canonical quantization of
a local field theory. The interaction dependence of the Poincar\'e
generators is then determined by the quantization surface. If the
surface is chosen as the forward hyperboloid $x_\mu x^\mu =
\tau^2$ all interactions go into $\hat{P}^\mu$ and one ends up
with point-form quantum field theory
(PFQFT)~\cite{Biernat:2007sz}. A quantized field theory is
intrinsically a many-body theory with infinitely many degrees of
freedom with the representation space being a Fock space.
Relativistic invariance follows automatically if one starts
 from a Poincar\'e-scalar Langrangean density.

The problem of constructing a relativistically invariant quantum
theory for a restricted number of particles is more intricate.
Interaction terms that show up in the dynamical Poincar\'e
generators are, in general, subject to non-linear constraints
which are imposed by the Poincar\'e algebra. A particular solution
to the problem of finding a consistent set of 10 Poincar\'e
generators for an interacting quantum mechanical system with
finitely many degrees of freedom was discovered by Bakamjian and
Thomas~\cite{Bakamjian:1953kh}. If their construction is carried
out in point form the resulting (interacting) 4-momentum operator
$\hat{P}^\mu$ is seen to separate into an interacting mass
operator $\hat{M}$ and a free velocity operator
$\hat{V}^\mu_{\mathrm{free}}$,
\begin{equation}
\hat{P}^\mu = \hat{P}^\mu_{\mathrm{free}} +
\hat{P}^\mu_{\mathrm{int}} =
\left(\hat{M}_{\mathrm{free}}+\hat{M}_{\mathrm{int}} \right)
\hat{V}^\mu_{\mathrm{free}}\, . \label{eq:PFBT}
\end{equation}
The PF Eqs.~(\ref{eq:pfeq}) imply that the interacting part of the
mass operator $\hat{M}_{\mathrm{int}}$ must be a Lorentz scalar
that  has to commute with the free velocity operator, i.e.
$[\hat{M}_{\mathrm{int}},\hat{V}^\mu_{\mathrm{free}}]=0$, to
ensure Poincar\'e invariance. Within the Bakamjian-Thomas
framework the operator for the total spin of the (interacting)
system is not affected by interactions. However, different forms
of dynamics are, in general, associated with different definitions
of the spin (canonical spin, helicity, front-form spin, $\dots$),
which are related via the unitary \lq\lq Melosh
transformation"~\cite{Keister:1991sb}. Remarkably, in the
Bakamjian-Thomas approach it is possible to construct Poincar\'e
invariant models with instantaneous interactions
(action-at-a-distance). What one looses with instantaneous
interactions as compared to a local quantum field theory is,
however, microcausality. In relativistic quantum mechanics this
property is replaced by the weaker condition of cluster
separability (or macrocausality)~\cite{Foldy:1960nb}.
Macrocausality roughly means that disjunct subsystems of a quantum
mechanical system should behave independently of each other, if
they are separated by sufficiently large space-like distances.

In Sec.~\ref{sec:PFQM} we will show that retardation effects in
particle-exchange interactions can be accommodated within the
Bakamjian-Thomas framework by means of a multichannel formulation.
This multichannel framework will then be used to derive
electroweak current matrix elements and form factors of bound
few-body systems. In this context it will be necessary to discuss
the question of cluster separability, as it plays a role in the
extraction of form factors. A further example shows that such a
coupled-channel approach could also be used to formulate
constituent quark models that lead to unstable excited states,
i.e. resonances with finite (non-perturbative) decay widths.
Section~\ref{sec:PFQFT} is devoted to point-form quantum field
theory. We will give a few applications for which field
quantization on a space-time hyperboloid could be advantageous as
compared to usual equal-time quantization.
Section~\ref{sec:conclusion} finally contains our conclusions and
an outlook.

\section{Point-form quantum mechanics}
\label{sec:PFQM}
\subsection{Relativistic multichannel framework}
Our first aim will be to describe particle-exchange interactions
within the Bakamjian-Thomas approach in such a way that
retardation effects are fully taken into account. This can be done
within a multichannel framework that allows for the creation and
annihilation of a finite number of additional particles. From
Eq.~(\ref{eq:PFBT}) it is clear that it suffices to study the
eigenvalue problem for the mass operator. Let us start with a
2-channel system in which an $N$-particle channel is coupled to an
$(N+1)$-particle channel. The mass-eigenvalue problem for such a
system can be written as
\begin{equation}
\left( \begin{array}{ll} \hat{M}_N & \hat{K}^\dag \\
\hat{K} & \hat{M}_{N+1}
\end{array}\right) \left( \begin{array}{l} \vert \Psi_N\rangle \\ \vert
\Psi_{N+1}\rangle
\end{array}\right)= m \left( \begin{array}{l} \vert \Psi_N\rangle \\ \vert
\Psi_{N+1}\rangle
\end{array}\right)\, , \label{eq:massev}
\end{equation}
with a matrix mass operator that acts on the direct sum of $N$-
and $(N+1)$-particle Hilbert spaces. $ \vert \Psi_N\rangle$ and
$\vert \Psi_{N+1}\rangle$ are the $N$- and $(N+1)$-particle
components of the mass eigenstate with mass eigenvalue $m$. The
diagonal elements of the mass operator $\hat{M}_n$, $n=N,N+1$
represent the invariant masses of the uncoupled $N$- and
$(N+1)$-particle systems. In addition to the relativistic kinetic
energies of the particles these operators may contain
instantaneous interactions between the particles. In the
forthcoming applications to constituent-quark models an
instantaneous confinement between the (anti)quarks is, e.g., put
into $\hat{M}_n$. $\hat{K}$ is the vertex operator that accounts
for the creation of the additional particle.

For our further purposes it is useful to apply a Feshbach
reduction to Eq.~(\ref{eq:massev}) and eliminate the
$(N+1)$-particle channel in favor of the $N$-particle channel:
\begin{equation}
m \, \vert \Psi_N\rangle = \left( \hat{M}_N + \hat{K}^\dag\,
(m-\hat{M}_{N+1})^{-1}\, \hat{K}\right) \vert \Psi_N\rangle =:
\Big( \hat{M}_N + \hat{V}_{\mathrm{opt}}(m) \Big) \vert
\Psi_N\rangle  \, . \label{eq:vopt}
\end{equation}
The optical potential $\hat{V}_{\mathrm{opt}}(m)$ introduced in
this way is a function of the mass eigenvalue $m$ and consists of
all possibilities to exchange the ($N+1$)st particle between the
remaining particles. It also includes loop contributions, i.e.
absorption by the emitting particle. $(m-\hat{M}_{N+1})^{-1}$
describes the propagation of the $(N+1)$-particle intermediate
state and is thus responsible for retardation effects.

In order to analyze Eq.~(\ref{eq:vopt}) in some more detail it is
appropriate to introduce a set of basis states that is tailored to
the point-form version of the Bakamjian-Thomas construction:
\begin{equation}
\vert v; \vec{k}_1, \mu_1; \dots; \vec{k}_n, \mu_n \rangle =
\hat{U}_{B_c(v)} \, \vert \vec{k}_1, \mu_1; \dots; \vec{k}_n,
\mu_n \rangle \quad {\mathrm{with}} \quad \sum_{i=1}^n \vec{k}_i =
0\, .
\end{equation}
Such states are called \lq\lq velocity states". A velocity state
is a usual multiparticle momentum state in the rest frame
($\sum_{j=1}^n \vec{k}_j = 0$) that is boosted to overall
4-velocity $v$ ($v_\mu v^\mu=1$) by means of a canonical spin
boost $B_c(v)$~\cite{Klink:1998zz}. The $\mu_j$s denote the spin
projections of the individual particles. A velocity state is a
simultaneous eigenstate of the free $n$-particle mass operator
$\hat{M}_{\mathrm{free},n}$ (eigenvalue $\mathcal{M}_n =
\sum_{i=1}^n \sqrt{m_i^2+\vec{k}_i^2}$) and the 4-velocity
operator $\hat{V}^\mu_{\mathrm{free},n}$ (eigenvalue $v^\mu$). In
a velocity-state representation of a Bakamjian-Thomas type model
the overall velocity of the system can always be factored out as a
velocity-conserving delta function, leaving an equation for the
pure internal motion. A further advantage of velocity states is
that angular momenta can be coupled together as in
non-relativistic quantum mechanics.

The velocity-state representation can now be used to specify the
vertex operator $\hat{K}$ by relating it to an appropriate field
theoretical interaction density $\hat{\mathcal{L}}_{\mathrm{int}}(x)$.
Since the total velocity of a Bakamjian-Thomas type system should
be conserved, one is led to define matrix elements of $\hat{K}$
via the relation~\cite{Klink:2000pp}
\begin{eqnarray}\label{eq:vertex}\lefteqn{
\langle v^\prime; \vec{k}^\prime_1, \mu^\prime_1; \dots;
\vec{k}^\prime_{n+1}, \mu^\prime_{n+1} \vert\hat{K} \vert v;
\vec{k}_1, \mu_1; \dots; \vec{k}_n, \mu_n \rangle}\\&&
\hspace{-0.2cm}=\mathcal{N}_{n+1,n} v^0
\delta^3(\vec{v}-\vec{v}^\prime)\langle \vec{k}^\prime_1,
\mu^\prime_1; \dots; \vec{k}^\prime_{n+1}, \mu^\prime_{n+1}
\vert\hat{\mathcal{L}}_{\mathrm{int}}(0)\, f(\Delta m)\,\vert
\vec{k}_1, \mu_1; \dots; \vec{k}_n, \mu_n \rangle\, , \nonumber
\end{eqnarray} with
$\mathcal{N}_{n+1,n}=(2\pi)^3/\sqrt{\mathcal{M}_{n+1}^\prime
\mathcal{M}_n}$ . $f(\Delta
m=|\mathcal{M}_{n+1}^\prime-\mathcal{M}_n|)$ denotes a vertex form
factor that one may introduce to compensate for missing
off-diagonal velocity contributions, as well as regulate
integrals. The prescription, Eq.~(\ref{eq:vertex}), preserves the
Lorentz structure of field theoretical vertex interactions, but it
violates locality of the vertex due to the {\em overall} velocity
conserving delta function $\delta(\vec{v}-\vec{v}^\prime)$ and the
kinematical factor $\mathcal{N}_{n+1,n}$. These quantities do not
only depend on the kinematics at the vertex, but also on the
kinematics of the spectator particles. Later on we will comment on
the consequences for macroscopic causality.

This relativistic coupled-channel framework with field-theory
motivated vertex interactions has already been applied to several
problems. In Ref.~\cite{Girlanda:2007ds} the mass operator for
$\pi$-$N$ and $N$-$N$ scattering has been derived using chiral
perturbation theory. In the following we are going to show how
electromagnetic hadron form factors can be extracted from the
optical one-photon-exchange potential for electron scattering off
confined (anti)quarks~\cite{Biernat:2009my}. A most recent
application, which we are also going to discuss, is the
construction of constituent quark models that lead to
resonance-like behavior for hadron excitations.

\subsection{Electromagnetic structure of hadrons}
Electromagnetic hadron form factors are observables that encode
the electromagnetic structure of hadrons. The theoretical analysis
of hadron form factors amounts to the question of how the
electromagnetic current $J^\mu$ of a hadron may be expressed in
terms of the electromagnetic currents of its constituents. It has long
been recognized that binding effects must appear in
$J^\mu$ and it cannot be a simple sum of constituent
currents~\cite{Siegert:1937yt}. The 3 basic constraints that a
model for the electromagnetic hadron current has to satisfy are
Poincar\'e covariance, current conservation and the requirement
that the hadron charge should not be renormalized by binding
effects (i.e. correct normalization of form factors at vanishing
momentum transfer)~\cite{Lev:1994au}. General procedures for the
construction of model currents of bound few-body systems that
satisfy these constraints are given in
Refs.~\cite{Lev:1994au,Klink:1998qf}. Both papers use the
point form of relativistic quantum mechanics and the Breit frame,
in which the constraints on $J^\mu$ are most easily satisfied, for
their analysis. Due to the kinematic nature of Lorentz boosts in
point form the model current constructed in the Breit frame is
then readily transformed into an arbitrary frame. Given the
bound-state wave function of a few-body system, this
construction generates a whole class of currents that satisfy the
3 basic constraints. One of the admissible currents is the
spectator current, in which only one of the constituents is
active, whereas the other ones are just spectators.

In the following we will present a different strategy and rather
try to {\em derive} a microscopic model for a hadron current that
is consistent with the binding forces than to construct it (cf.
Ref.~\cite{Biernat:2009my}). To do this we will use the Poincar\'e
invariant coupled channel framework just introduced and study the
full physical electron-hadron scattering process and not just the
emission and absorption of a virtual photon by the hadron. To be
more specific, let us consider a pion within a constituent-quark
model. Electron-pion scattering is then treated as a 2-channel
problem. The first channel is the $e$$q$$\bar{q}$ channel, the
second the $e$$q$$\bar{q}$$\gamma$ one. In this way the dynamics
of the exchange photon is fully taken into account. The diagonal
terms of the mass operator in Eq.~(\ref{eq:massev}) are given by
$\hat{M}_3 = \hat{M}_{\mathrm{free},
eq\bar{q}}+\hat{V}_{\mathrm{conf}}$ and $\hat{M}_4 =
\hat{M}_{\mathrm{free}, eq\bar{q}\gamma}+\hat{V}_{\mathrm{conf}}$,
respectively. $\hat{V}_{\mathrm{conf}}$ is an {\em instantaneous}
confinement potential that acts only between the quark and
antiquark. The vertex operator $\hat{K}$ is defined via
velocity-state matrix elements as in Eq.~(\ref{eq:vertex}) with
$\hat{\mathcal{L}}_\mathrm{int}(x)$ being the Lagrangean density
for the coupling of a photon to a spin-1/2 (anti)quark.

Hadron form factors can be extracted from the invariant
1-photon-exchange amplitude for electron-hadron scattering. In our
case this amplitude is given by on-shell matrix elements of (the
1-photon-exchange part of) the optical potential between velocity
eigenstates of $\hat{M}_3$ with the $q$$\bar{q}$ cluster
possessing the quantum numbers of the pion. As one would expect,
these on-shell matrix elements of the optical potential can be
written as a contraction of the usual point-like electron current
with a hadronic current times the (covariant) photon propagator:
\begin{eqnarray}\label{eq:contract}\lefteqn{
\langle v^\prime; \vec{k}_e^\prime, \mu_e^\prime;
\vec{k}_M^\prime, \alpha_M \vert \hat{V}_{\mathrm{opt}}(m) \vert
v; \vec{k}_e, \mu_e; \vec{k}_M,
\alpha_M\rangle_{\mathrm{on-shell}}}\nonumber\\ &&
\hspace{3cm}\propto v^0 \delta^3(\vec{v}-\vec{v}^\prime)\frac{
j_\mu(\vec{k}_e^\prime, \mu_e^\prime; \vec{k}_e, \mu_e)
J^\mu(\vec{k}_M^\prime;\vec{k}_M)}{(k_e^\prime-k_e)^2}\, .
\end{eqnarray}
\lq\lq On-shell" means that $m=k_e^0+k_M^0=k_e^{\prime
\,0}+k_M^{\prime\, 0}$ and $k_e^0=k_e^{\prime \,0}$ and
$k_M^0=k_M^{\prime \,0}$, $\alpha_M$ denotes the discrete quantum
numbers of $q$$\bar{q}$ bound state. The detailed derivation of
Eq.~(\ref{eq:contract}) and the explicit expression for the pion
current $J^\mu(\vec{k}_M^\prime;\vec{k}_M)$ can be found in
Ref.~\cite{Biernat:2009my}. Since $\vec{k}_M$ and
$\vec{k}_M^\prime$ are momenta defined in the center of mass of
the electron-meson system $J^\mu(\vec{k}_M^\prime;\vec{k}_M)$ does
not behave like a 4-vector under a Lorentz transformation
$\Lambda$. It rather transforms by the Wigner rotation
$R_{\mathrm{W}}(v,\Lambda)$. The current with the correct
transformation properties is the one involving the physical meson
momenta $p_M^{(\prime)}=B_c(v)\,k_M^{(\prime)}$:
\begin{equation}\label{eq:mcurrent}
J^\mu(\vec{p}_M^\prime;\vec{p}_M) :=
[B_c(v)]^\mu_{\phantom{\mu}\nu}\,
J^\nu(\vec{k}_M^\prime;\vec{k}_M) \, .
\end{equation}
In addition one can show that $(p_M^\prime - p_M)_\mu\,
J^\mu(\vec{p}_M^\prime;\vec{p}_M)=0$, i.e. $J^\mu$ is a conserved
current. If $J^\mu(\vec{p}_M^\prime;\vec{p}_M)$ were a correct
model for the pion current it would be possible to write it in
the form $J^\mu(\vec{p}_M^\prime;\vec{p}_M) = (p_M+p_M^\prime)^\mu
F(Q^2)$ (with $Q^2=\vert (p_M^\prime-p_M)^2 \vert$) which should
hold for arbitrary values of the meson momenta $p_M$ and
$p_M^\prime$. This is, however, not possible in our case. The
reason is that our derivation of the current is based on the
Bakamjian-Thomas construction which is known to provide wrong
cluster properties~\cite{Keister:1991sb}. Therefore we cannot be
sure that the hadronic current we get does not also depend on the
electron momenta. What we find is indeed that
$J^\mu(\vec{p}_M^\prime;\vec{p}_M)$ cannot be fully expressed in
terms of hadronic covariants, but one needs 1 additional (current
conserving) covariant, which is the sum of the 2 electron momenta:
\begin{equation}\label{eq:Jparam}
J^\mu(\vec{p}_M^\prime;\vec{p}_M) = (p_M+p_M^\prime)^\mu f_1(Q^2,
s)+(p_e+p_e^\prime)^\mu f_2(Q^2, s)\, .
\end{equation}
This decomposition holds for arbitrary values of the meson momenta
$p_M$ and $p_M^\prime$ with only one exception, the Breit frame
(i.e. backward scattering in the electron-meson center of mass
system). In this frame the 2 covariants become proportional and it
is not possible to separate the 2 form factors. Wrong cluster
properties may also influence the form factors $f_i$ such that
they not only depend on the squared 4-momentum transfer at the
photon-pion vertex, but also on Mandelstam $s$, i.e. the square of
the invariant mass of the electron-meson system.
\begin{figure}
\includegraphics[width=6.0cm]{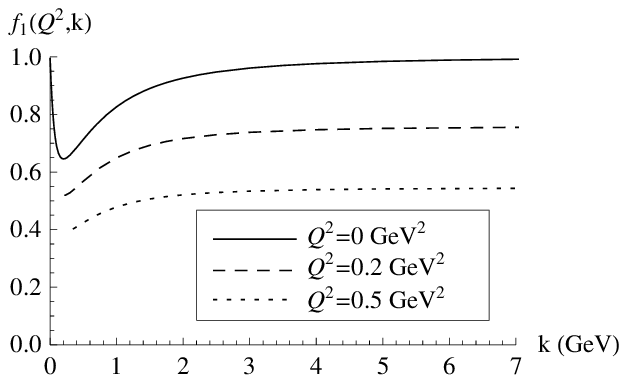}
\includegraphics[width=6.0cm]{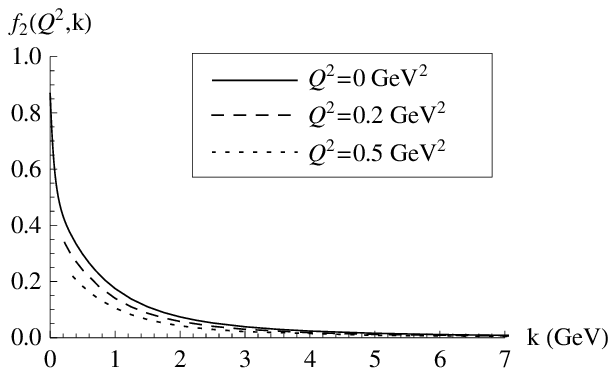}
\caption{Dependence of the physical pion form factor $f_1$ and of
the spurious pion form factor $f_2$ (cf. Eq.~(\ref{eq:Jparam})) on
the meson center-of-mass momentum $k=|\vec{k}_M|$ for different
values of the the squared momentum transfer $Q^2$.}\label{fig:ffs}
\end{figure}

The influence of cluster-separability violating effects can be
studied numerically. To this end we have taken the simple harmonic
oscillator wave function for the $q$$\bar{q}$ bound state that has
already been used in Ref.~\cite{Biernat:2009my} to calculate the
microscopic current $J^\mu(\vec{k}_M^\prime;\vec{k}_M)$ and
separate the form factors by means of Eq.~(\ref{eq:Jparam}). The
result is shown in Fig.~\ref{fig:ffs}. One first observes that
both, the physical form factor $f_1$ and the spurious form factor
$f_2$ depend on $|\vec{k}_M|$ (or Mandelstam $s$). This $s$
dependence of $f_1$ vanishes rather quickly with increasing invariant
mass. At the same time the spurious form factor $f_2$ is seen to
vanish. It is thus suggestive to take the limit $s\rightarrow
\infty$ to get a sensible result for the pion form factor. In this
limit the microscopic pion current factorizes explicitly into a
point-like pion current times a $Q^2$ dependent integral which can
be identified as pion form factor:
\begin{equation}
J^\mu(\vec{k}_M^\prime;\vec{k}_M) \stackrel{s\rightarrow
\infty}{\longrightarrow} (k_M+k_M^\prime)^\mu F(Q^2) \, .
\end{equation}
The final result for the form factor has a simple analytical form
and can be proved to be equivalent to the standard front form
expression for a spectator current in the $q^+=0$
frame~\cite{Biernat:2009my}. The fact that the $s\rightarrow
\infty$ limit removes effects due to violation of cluster
separability seems to have a plausible explanation: In order to
separate subsystems by space-like distances in point form it
is natural to use boosts. In this way one stays on the
quantization hypersurface. Separation by boosts, however, increases
the invariant mass of the whole system. Or, reversing the
argument, by taking $s$ large we separate the electron from the
meson such that it does not affect the meson structure.

This procedure for the calculation of form factors can immediately
be generalized to electroweak form factors of arbitrary bound
few-body systems. Very recently it has been applied to electroweak
form factors of heavy-light mesons~\cite{Gomez:2010}. We have been
able to derive a simple analytical expression for the Isgur-Wise
function and to prove that the form-factor relations due to
heavy-quark symmetry are recovered in the heavy-quark limit
$m_Q\rightarrow \infty$. The electromagnetic form factors of
spin-1 bound states, like the deuteron, are presently under
investigation. Here it turns out that one cannot completely get
rid of the problems associated with cluster separability by taking
the $s\rightarrow \infty$ limit. Covariant structures in which
$(k_e+k_e^\prime)$ is contracted with the in- and outgoing
polarization vectors of the spin-1 particle do not vanish in this
limit. But our findings resemble those of
Ref.~\cite{Carbonell:1998rj} that have been obtained within a
covariant light-front approach. In this paper the authors also
encounter spurious contributions to the current which are
associated with a 4-vector $\omega^\mu$ that characterizes the
orientation of the light front. These spurious contributions
correspond exactly to our spurious contributions which have their
origin in the violation of cluster separability. Their presence
guarantees that the angular condition for the (spin) matrix
elements of the physical part of the current is
satisfied~\cite{Carbonell:1998rj}. This means that there is also a
connection between the angular condition and cluster separability.


%
\subsection{Dressing of hadrons and decays of resonances}
The resonance character of hadron excitations is usually ignored
when calculating hadron spectra within constituent quark models.
Hadron excitations rather come out as stable bound states and
their bound-state wave function is subsequently used to calculate
partial decay widths perturbatively by assuming a particular model
for the decay vertex on the constituent level. The observation
that the predicted strong decay widths are notoriously too
small~\cite{Melde:2006yw,Sengl:2007yq,Metsch:2008zz} is an
indication that a physical hadron resonance is not just a simple
bound state of valence (anti)quarks, but it should also contain
(anti)quark-meson components. A natural starting point for
including such components is the chiral constituent-quark model.
The physical picture behind this model is that the effective
degrees of freedom emerging from chiral symmetry breaking of QCD
are constituent quarks and Goldstone bosons which couple directly
to the constituent quarks. The Goldstone bosons can be identified
with the lightest pseudoscalar meson octet and singlet. With a
linear confinement and the hyperfine interaction being given by
the instantaneous approximation to Goldstone boson exchange the
chiral constituent-quark model has been very successful in
reproducing mass spectra~\cite{Glozman:1997ag} and electroweak
form factors of light baryons~\cite{Boffi:2001zb}. These
calculations (and also those in
Refs.~\cite{Melde:2006yw,Sengl:2007yq}) were already done within
the framework of relativistic point-form quantum mechanics.

\begin{figure}
\begin{center}
\includegraphics[height=3.0cm]{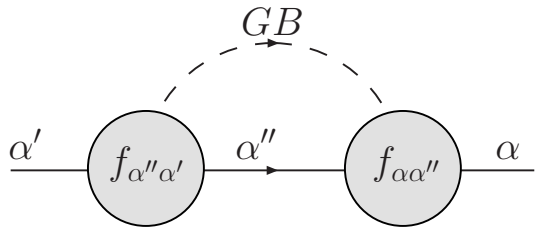}
\end{center} \caption{Graphical representation of the optical potential,
Eq.~(\ref{eq:vopth}), that enters the mass eigenvalue
equation~(\ref{eq:massevh}) on the hadronic level.}
\label{fig:vopt}
\end{figure}

We will now go one step further and take the dynamics of the
Goldstone-boson exchange fully into account by employing the
relativistic coupled channel framework introduced at the beginning
of this section. If we want to study baryon resonances, we have in
the first channel 3 quarks and in the second channel 3 quarks plus
a Goldstone boson. A physical mass eigenstate is then a
superposition of a 3-quark component $\vert \Psi_{3} \rangle$ and
a 3-quark plus Goldstone-boson component $\vert \Psi_{4}\rangle$.
These components are solutions of the mass eigenvalue
equation~(\ref{eq:massev}) with the diagonal terms of the mass
operator being given by $\hat{M}_3 = \hat{M}_{\mathrm{free},
3q}+\hat{V}_{\mathrm{conf}}$ and $\hat{M}_4 =
\hat{M}_{\mathrm{free}, 3q+GB}+\hat{V}_{\mathrm{conf}}$,
respectively. $\hat{V}_{\mathrm{conf}}$ is an {\em instantaneous}
confinement potential that acts only between the quarks. The
vertex operator $\hat{K}$ is defined via velocity-state matrix
elements as in Eq.~(\ref{eq:vertex}) with $\hat \mathcal{L}_\mathrm{int}(x)$ being
the Lagrangean density for the coupling of a pseudoscalar particle
to a spin-1/2 particle. The physical picture that emerges from
this model becomes most obvious if we study, instead of
Eq.~(\ref{eq:massev}), the equivalent (non-linear) eigenvalue
equation~(\ref{eq:vopt}). In order to convert it into an algebraic
equation we expand $\vert \Psi_{3} \rangle$ in terms of (velocity)
eigenstates $\vert v; \alpha \rangle$ of $\hat{M}_3$, i.e.
eigenstates of the pure confinement problem:
\begin{equation}
\vert \Psi_3 \rangle = \sum_\alpha \int \frac{\mathrm{d}^3v}{(2
\pi)^3} m_\alpha^2 \delta^3(\vec{v}-\vec{V}) A_\alpha \vert v;
\alpha \rangle \, .
\end{equation}
Here, $\vec{V}$ denotes the overall velocity of the coupled
$3q$-$GB$ system and $\alpha$ the whole set of discrete quantum
numbers that are necessary to specify a particular eigenstate of
the pure confinement problem uniquely. $\vec{v}$ is the velocity
of such an eigenstate and $m_\alpha$ its mass. For reasons which
will become clear very soon we will call $\vert v; \alpha \rangle$
a \lq\lq bare" baryon state and $\vert \Psi_3\rangle$ the (3-quark
component of the) \lq\lq physical" baryon state. By projecting
Eq.~(\ref{eq:vopt}) onto bare baryon states $\vert v; \alpha
\rangle$ and inserting the completeness relation for such states
between the operators and $\vert \Psi_3\rangle$ we end up with an
infinite set of algebraic equations for the expansion coefficients
$A_\alpha$:
\begin{eqnarray}\label{eq:massevh}
2v^0 \delta^3(\vec{v}-\vec{V}) (m_\alpha - m) A_\alpha =
\sum_{\alpha^\prime} \int \frac{\mathrm{d}^3v}{(2 \pi)^3} \,
m_{\alpha^\prime}^2 \, \langle v; \alpha \vert
\hat{V}_{\mathrm{opt}}(m) \vert v^{\prime} ; \alpha^\prime\rangle
\, \delta^3(\vec{v}^\prime -\vec{V})\, A_{\alpha^\prime}\, .\nonumber\\
\end{eqnarray}
This equation is now a mass-eigenvalue equation for hadrons rather
than for quarks. It describes how a physical hadron of mass $m$ is
composed of bare hadrons with masses $m_\alpha$. The bare hadrons
are mixed via the optical potential $\hat{V}_{\mathrm{opt}}(m)$
which we will now analyze in some more detail. By inserting
completeness relations for eigenstates of $\hat{M}_3$,
$\hat{M}_4$, $\hat{M}_{\mathrm{free}, 3q}$, and
$\hat{M}_{\mathrm{free}, 3q+GB}$ at appropriate places one infers
that the optical potential has the following structure:
\begin{eqnarray}\label{eq:vopth}
\langle v; \alpha \vert \hat{V}_{\mathrm{opt}}(m) \vert v^{\prime}
; \alpha^\prime\rangle &=& g^2 \sum_{\alpha^{\prime\prime}} v^0
\delta^3(\vec{v}-\vec{v}^\prime) \int \frac{\mathrm{d}^3 \kappa}{2
\sqrt{m_{GB}^2 + \vec{\kappa}^2}} \frac{1}{2
\sqrt{m_{\alpha^{\prime\prime}}^2+\vec{\kappa}^2}}\nonumber\\
&&\hspace{-0.5cm}\times \frac{1}{\sqrt{m_{\alpha^\prime}^3
m_\alpha^3}}\, \frac{f_{\alpha \alpha^{\prime\prime}}(\vert
\vec{\kappa}\vert) \, f_{\alpha^{\prime\prime}
\alpha^{\prime}}(\vert
\vec{\kappa}\vert)}{m-\sqrt{m_{\alpha^{\prime\prime}}^2+\vec{\kappa}^2}-\sqrt{m_{GB}^2
+ \vec{\kappa}^2}+i\epsilon}\, .
\end{eqnarray}
This means that the optical potential couples bare baryon states
$\vert v^{\prime} ; \alpha^\prime\rangle$ and $\vert v ; \alpha
\rangle$ via a Goldstone boson loop such that any bare baryon
state $\vert v^{\prime\prime} ; \alpha^{\prime\prime}\rangle$
(that is allowed by conservation laws) can be excited in an
intermediate step (cf. Fig.~\ref{fig:vopt}). $g$ is the
quark-Goldstone boson coupling, $f_{\alpha \alpha^{\prime}}(\vert
\vec{\kappa}\vert)$ are (strong) transition form factors that show
up at the (bare) baryon Goldstone-boson vertices. With this
expression for the matrix elements of the optical potential the
velocity dependence in Eq.~(\ref{eq:massevh}) is seen to cancel
out completely. The eigenvalue problem that one ends up with
describes thus bare baryons, i.e. eigenstates of the pure
confinement problem, that are mixed and dressed via
Goldstone-boson loops. The only places where the quark
substructure enters, are the vertex form factors. Here we want to
emphasize that due to the instantaneous nature of the confinement
potential the dressing happens on the baryonic level and not on
the quark level, i.e. emission and absorption of the Goldstone
boson by the same quark must not be interpreted as mass
renormalization of the quark. This was done, e.g., in
Ref.~\cite{Krassnigg:2003gh} where this kind of coupled channel
approach has been applied for the first time to calculate the mass
spectrum of vector mesons within the chiral constituent quark
model.

Note that the optical potential, Eq.~(\ref{eq:vopth}), becomes
complex if the mass eigenvalue $m$ is larger than the lowest
threshold $m_{\mathrm{th}}=m_0+m_{GB}$, i.e. the mass of the
lightest bare baryon plus the Goldstone-boson mass. As a
consequence also the physical mass eigenvalues $m$ will become
complex as soon as their real part is larger than
$m_{\mathrm{th}}$ and we will get unstable baryon excitations. The
mass of such an excitation can then be identified with
$\mathrm{Re}(m)$, its width $\Gamma$ with $2\, \mathrm{Im}(m)$.
The mass eigenvalue problem, Eq.~(\ref{eq:massevh}), can be solved
by an iterative procedure. One first has to restrict the number of
bare states, that are taken into account, to a certain number
$\alpha_{\mathrm{max}}$. The first step is to insert a start value
for $m$ into $\hat{V}_{\mathrm{opt}}(m)$ and solve the resulting
linear eigenvalue equation. This leads to $\alpha_{\mathrm{max}}$
(possibly complex) eigenvalues. From these one has to pick out the
right one, reinsert it into $\hat{V}_{\mathrm{opt}}(m)$, solve
again, etc. Appropriate start values are, e.g., the eigenvalues of
the pure confinement problem. In order to pick out the correct
eigenvalue in each step it is helpful to perform this procedure
by gradually increasing the strength of the optical potential (until
it has reached its full strength) and observe the path of the
eigenvalues in the complex plane.

A first numerical test of the ideas just presented has been
performed for a toy model in Ref.~\cite{Kleinhappel:2010}. The toy
model can be understood as a simplified constituent-quark model
for mesons with its degrees of freedom being a quark, an antiquark
and a real scalar meson which couples to the (anti)quark. Both,
quark and antiquark are treated as spin- and flavorless particles
that are confined by an instantaneous harmonic-oscillator
potential. To simplify things further, only bare hadron states
that correspond to radial (and not orbital) excitations of the
pure confinement problem are taken into account when solving
Eq.~(\ref{eq:massevh}). Reasonable convergence for the ground
state and the first excited state is already achieved with
$\alpha_{\mathrm{max}}= 4$ after 5 iterations. With a meson-quark
coupling compatible with the Goldberger-Treiman relation and a
reasonable parameterization of the confinement potential the 2
lowest lying states are found to have masses of about $0.8$ and
$1.45$~GeV, respectively. The first excited state has a width of
$0.026$~GeV. These are promising results in view of the simplicity
of the model and it will be interesting to see whether typical
decay widths of $0.1$~GeV or more can be achieved in the case of
baryon resonances with the full chiral constituent-quark model.

The picture that emerges from the chiral constituent-quark model
with instantaneous confinement and dynamical Goldstone-boson
exchange corresponds to the kind of hadronic resonance model that
has been developed and refined by Sato and Lee~\cite{Sato:2009de}
and that is now heavily applied at the Excited Baryon Analysis
Center (EBAC) at JLab to extract $N^\ast$ properties from the
world data on $\pi N$, $\gamma N$, and $N(e,e^\prime)$ reactions.
Admittedly, our attempts are still far away from the degree of
sophistication of the Sato-Lee model, but unlike this pure
hadronic model we could also predict the meson-baryon couplings
and vertex form factors from the underlying constituent-quark
model. Our approach is in a certain sense inverse to the Sato-Lee
approach. Whereas they want to \lq\lq undress" physical resonances
to end up with \lq\lq bare" quantities that can, e.g., be compared
with results from naive constituent quark models, we rather want
to \lq\lq dress" constituent-quark models to end up with physical
quantities that can be directly compared with experiment.

\section{Point-Form Quantum Field Theory}
We now turn to infinite degree-of-freedom systems satisfying the
point-form equations.  Irreducible representations of the
Poincar\'e group generate one-particle Hilbert spaces for massive
and massless particles.  Infinite tensor products of these
one-particle Hilbert spaces then generate many-body Fock spaces.
Alternatively creation and annihilation operators with the correct
Poincar\'e transformation properties form an algebra of operators,
in which the Fock space is generated by the action of products of
creation operators acting on the Fock vacuum.  Further both the
free and interacting four-momentum operators are polynomials in
the creation and annihilation operators. \label{sec:PFQFT}

\subsection{Free quantum fields}
Consider first massive particles of spin j.  The single-particle
Hilbert space is then $L^2(SO(1,3)/SO(3))\times V^j$, the space of
square-integrable functions over the forward hyperboloid times a
2j+1 dimensional spin space.  On this space the natural variable
is the four-velocity, $v:=p/m$, the four-momentum divided by mass.
Under a Poincar\'e transformation, a one-particle state of
four-velocity $v$ and spin projection $\sigma$ transforms as
\begin{eqnarray}
\hat{U}_a\vert v,\sigma\rangle&=&e^{ip\cdot a} \vert v,\sigma\rangle\, ,\\
\hat{U}_{\Lambda}\vert v,\sigma\rangle&=&\sum \vert\Lambda
v,\sigma'\rangle D^j_{\sigma'\sigma} (R_W(v,\Lambda)),
\end{eqnarray}
where $a$ is a four-translation and $D^j_{\sigma'\sigma}(\ldots)$
 an $SU(2)$ matrix element for spin $j$.  $R_W(v,\Lambda)$
is a Wigner rotation, an element of the rotation group $SO(3)$
defined by
\begin{eqnarray}
R_\mathrm W(v,\Lambda):&=&B_c^{-1}(\Lambda v)\Lambda B_c(v).
\end{eqnarray}
$B_c(v)$ is a boost, a Lorentz transformation satisfying
$v=B_c(v)v^{\mathrm {rest}}$, with $v^{\mathrm{rest}}=(1,0,0,0)$.
Here boosts are canonical spin boosts.

 Classical irreducible fields over Minkowski space-time, with a differentiation
 inner product, are then naturally related to the one-particle wave
 functions over the four-velocity. As an example, consider a scalar particle
 with wave function
$\phi(v)\in L^2(R^3)$:
\begin{eqnarray}
||\phi||^2&=&\int \mathrm dv |\phi(v)|^2; \quad \mathrm dv:=\frac{\mathrm d^3 v}{(2\pi)^{3/2}2v_0}\\
\psi(x)&=&\int \mathrm dv\, e^{-imv\cdot x}\phi(v)\\
||\psi||^2&=&\int \mathrm dx^{\mu}
\left(\frac{\partial \psi^{\ast}}{\partial x^{\mu}}\psi-\psi^{\ast}
\frac{\partial \psi}{\partial x^{\mu}}\right)\nonumber\\
&=&||\phi||^2.
\end{eqnarray}
Note that the integration on space-time $\mathrm dx^{\mu}:= \mathrm d^4
x\,\delta(x\cdot x-\tau^2)\theta(x_0)x^{\mu}$ is over the forward
hyperboloid, typical of the point form.

To generate a many-particle theory introduce creation and
annihilation operators with the usual commutation relations for
fermions ($\hat a(v,\sigma,\alpha)$), antifermions ($\hat b(v,\sigma,\alpha)$) and
bosons ($\hat c(v,\lambda, i)$). Products of bilinears $\hat a^{\dagger}\hat a,
\hat a^{\dagger}\hat  b^{\dagger},\hat b\hat a,\hat b\hat b^{\dagger}$ along with products of
boson creation and annihilation operators form the algebra of
operators $\mathcal{A}$, which, acting on the Fock vacuum,
generates the usual Fock space.

The free four-momentum operator is made from this algebra of
operators:
\begin{eqnarray}
\hat P^{\mu}_{\mathrm{free}}:&=&m\sum \int \mathrm dv\,
v^{\mu}\left(\hat a^{\dagger}(v,\sigma,\alpha)\hat a(v,\sigma,\alpha)\right.\nonumber\\
&& \left.-\hat b(v,\sigma,\alpha) \hat b^{\dagger}(v,\sigma,\alpha)\right)+\kappa
\left(\hat c^{\dagger}(v,\lambda,i)\hat c(v,\lambda,i)\right),
\end{eqnarray}
where  $\kappa$ is a dimensionless relative bare boson mass
parameter and $m$ is a constant with the dimensions of mass.
$\alpha$ and $i$ are internal symmetry variables. It is easily
checked that the free four-momentum operator satisfies the point-form equations.

To generate quantum fields for massive particles, it is
straightforward to use the mapping between irreps of the
Poincar\'e group and classical irreducible fields to define (for
example spin 1/2 and 0)
\begin{eqnarray}
\hat \Psi_{\alpha}(x)&=&\sum\int  \mathrm dv(\hat a(v,\sigma,\alpha)u_{\sigma}(v)e^{-imv\cdot x}+
\hat b^{\dagger}(v,\sigma,\alpha)v_{\sigma}(v)e^{imv\cdot x})\, ,\\
\hat \phi_{i}(x)&=&\int \mathrm dv\left(\hat c(v, i)e^{-imv\cdot x}+\hat c^{\dagger}(v,
i)e^{imv\cdot x}\right).
\end{eqnarray}

It is then possible to define the usual free Lagrangean and get
$\hat P^{\mu}_{\mathrm{free}}$ from the stress energy tensor,
\begin{eqnarray}
\hat P^{\mu}_{\mathrm{free}}&=&\int\mathrm d^4 x\,\delta(x\cdot x-\tau^2)\theta(x_0)x_{\nu}\hat T^{\mu\nu}_{\mathrm{free}}\nonumber\\
&=&\int \mathrm dx_\nu \hat T^{\mu\nu}_{\mathrm{free}}.
\end{eqnarray}
As an example, consider the free charged scalar field,
\begin{eqnarray}
\hat{\mathcal{L}}&=&\partial^{\mu}\hat \phi^{\dagger}(x)\partial_{\mu}\hat \phi(x)-
m^2\hat \phi^{\dagger}(x)\hat \phi(x);\\
\hat T^{\mu\nu}&=&\partial^{\mu}\hat \phi^{\dagger}(x)\partial^{\nu}\hat \phi(x)+
\partial^{\nu}\hat \phi^{\dagger}(x)\partial^{\mu}\hat \phi(x)-\eta^{\mu\nu}
\hat \mathcal{L}.
\end{eqnarray}
As shown in Ref.~\cite{Biernat:2007sz} the free four-momentum
operator obtained from such a Lagrangean is the same as the one
obtained from irreps of the Poincar\'e group.

The same sort of analysis can be carried out for massless
particles. However, now the one-particle Hilbert space is given by
square integrable functions over the forward light cone times a
spin space, where the light cone is given by $SO(1,3)/E(2)$, the
homogeneous space defined by the ratio of the Lorentz group
divided by the two dimensional Euclidean group.  In the following
we will consider only gluon representations, as that is what is
relevant for point-form QCD. In that case the spin space can be
chosen to be the four dimensional nonunitary irrep of E(2) to get
four polarization degrees of freedom (labeled $\rho$). The
standard four-vector $k^{\mathrm{st}}=(1,0,0,1)$ leaves E(2)
invariant; a helicity boost,
$B(k):=R(\hat{k})\Lambda_z(|\vec{k}|)$, to a four-momentum $k$
then generates a gluon state with transformation properties
\begin{eqnarray}
\hat U_{e_2}\vert k^{\mathrm{st}}, \rho, a\rangle&=&\sum\vert
k^{\mathrm {st}},
\rho',a\rangle\Lambda_{\rho'\rho}(e_2),\\
\vert k,\rho, a\rangle:&=&\hat U_{B(k)}\vert k^{\mathrm{st}},\rho, a\rangle,\\
\hat U_\Lambda\vert k,\rho ,a\rangle&=&\hat U_\Lambda \hat U_{B(k)}\vert k^{\mathrm {st}},\rho, a\rangle\nonumber\\
&=&\sum\vert \Lambda k,\rho',a\rangle\Lambda_{\rho'\rho}(e_\mathrm W),\\
\hat U_c\vert k,\rho ,a\rangle&=&\sum \vert k,\rho ,a'\rangle D_{a'a}(c),
\end{eqnarray}
where $\Lambda(e_\mathrm W)=B^{-1}(\Lambda k)\Lambda B(k)$  is a Euclidean
Wigner "rotation", $c$ is an element of the internal symmetry
(color) group and $a,a'$ are color indices.

A Lorentz invariant gluon inner product is given by
 \begin{eqnarray}
 ||\phi||^2=-\sum\int \mathrm dk\, \eta_{\rho\rho}|\phi(k,\rho,a)|^2,
 \end{eqnarray}
 and shows the origin of gauge degrees of freedom, for this inner product
 is not positive definite.  By imposing a
Lorentz invariant auxiliary condition $\phi(k,0,a)=\phi(k,3,a)$,
the inner product becomes positive definite, and then only two
polarizations are physical, as required for massless particles.

A classical gluon field is  defined by
\begin{eqnarray*}
G^{\mu}_{a}(x)=\int\mathrm dk\, B^{\mu\rho}(k)\phi(k,\rho,a)e^{-ik\cdot x},
\end{eqnarray*}
with a norm given by a differentiation inner product over the
forward hyperboloid.  It should be noted that the polarization
matrix is a Lorentz boost.  Further it can be shown that
$||G||^2=||\phi||^2$.

As with massive particles, many-gluon states are generated by
gluon creation and annihilation operators such that
 \begin{eqnarray}
 \vert k,\rho,a\rangle&=&\hat g^\dagger(k,\rho,a)\vert 0\rangle\, ,\\
 \hat g(k,\rho,a)\vert 0\rangle&=&0,\quad \forall k,\rho,a\, ,\\
  {[\hat g(k,\rho,a),\hat g^\dagger(k',\rho',a')]}&=&-\eta_{\rho\rho'} k_0
\delta^3(\vec k -\vec k')\delta_{aa'}\, ,\\
 \hat U_\Lambda\, \hat g(k,\rho,a)\,\hat U_\Lambda^{-1}&=&\sum \hat
 g(\Lambda,k,\rho',a)\Lambda_{\rho\rho'}(e_\mathrm W)\, ,\\
 \hat U_c \,\hat g(k,\rho,a)\,\hat U_c^{-1}&=&\hat g(k,\rho,a')D_{a'a}(c)\, ,\\
 \hat P^{\mu}_{\mathrm{free}}&=&-\sum\int \frac{\mathrm d^3k}{k_0}k^\mu
\hat g^\dagger(k,\rho,a)\eta^{\rho\rho} \hat g(k,\rho,a).
 \end{eqnarray}
The auxiliary condition eliminating the 0 and 3 components on the
physical many-body Fock space is the annihilation operator
condition,
\begin{eqnarray}
\sum k^{\mathrm{st}}_{\rho}\eta^{\rho\rho}\hat g(k,\rho,a)\vert\phi\rangle=0,
\end{eqnarray}
which is Lorentz invariant.

Gauge transformations, which change the 0 and 3 components only,
are given by
\begin{eqnarray}
\hat g(k,\rho, a)\rightarrow \hat g'(k,\rho,a)&=&\hat g(k,\rho,a)+k^{\mathrm{st}}_{\rho} f(k,a)\hat I\\
\hat g^{\dagger}(k,\rho,a)\rightarrow
\hat g^{\prime\dagger}(k,\rho,a)&=&\hat g^{\dagger}(k,\rho,a),
\end{eqnarray}
and leave the  commutation relations and auxiliary condition
unchanged.

The free gluon field is again defined through the gluon creation
and annihilation operators:
\begin{eqnarray}
\hat G^{\mu}_a(x)&=&\int \mathrm dkB^{\mu\rho}(k)\left(e^{-ik\cdot x}\hat g(k,\rho, a)+
e^{ik\cdot x}\hat g^{\dagger}(k,\rho ,a)\right)\, ,\\
\frac{\partial}{\partial x^\nu} \frac{\partial}{\partial x_\nu}
\hat G^\mu_a&=&0.
\end{eqnarray}
Under a gauge transformation this gives
\begin{eqnarray}
\hat G^\mu_a (x)\rightarrow \hat G^{\prime\mu}_a (x)&=&\hat G^\mu_a (x)+\frac{\partial
\tilde{f}(x,a)}{\partial x_\mu}\hat I;\\
 \tilde{f}(x,a)&=&\int \frac{\mathrm d^3 k}{2k_0} e^{-ik\cdot x}f(k,a),
\end{eqnarray}
which preserves the Lorentz gauge on the physical Fock space:
\begin{eqnarray}
\partial \hat G^{ +}_{\mu,a} (x)/\partial x_\mu\vert\phi\rangle&=&i\int \mathrm dk\, k^\mu
B_{\mu\rho}(k) \eta_{\rho,\rho}e^{-ik\cdot x} \hat g(k,\rho,a)\vert\phi\rangle\nonumber\\
&=&i\sum \int\mathrm dk\, e^{-ik\cdot x}\,k^{\mathrm{st}}_\rho \eta_{\rho\rho}
\hat g(k,\rho, a)\vert \phi\rangle\nonumber\\
&=&0.
\end{eqnarray}

The Lagrangean for free gluon fields generates a stress energy
tensor and a free gluon four-momentum operator, which agrees with
the four-momentum operator given by gluon creation and
annihilation operators.  Rather than writing this out, we  next
turn to interacting quantum fields and give as an example the
self-coupling interaction of gluon fields.

\subsection{Interacting quantum fields}
Interactions in PFQFT are generated by
integrating vertices over the forward hyperboloid.
 A vertex $\hat \mathcal {L}_{\mathrm{int}}(x)$ is a local space-time scalar density operator,
\begin{eqnarray}
e^{-i\hat P_{\mathrm {free}}\cdot a} \hat \mathcal {L}_{\mathrm{int}}(x)e^{i
\hat P_{\mathrm {free}}\cdot a}&=& \hat \mathcal {L}_{\mathrm{int}}(x+a)\, ,\\
\hat U_{\Lambda}\hat \mathcal {L}_{\mathrm{int}}(x)\hat U_{\Lambda}^{-1}
&=&\hat \mathcal {L}_{\mathrm{int}}(\Lambda x)\, ,\\
\left[\hat \mathcal {L}_{\mathrm{int}}(x),\hat
\mathcal {L}_{\mathrm{int}}(y)\right]&=&0,\quad (x-y)^2 \,\mathrm {spacelike}.
\end{eqnarray}

Then the interacting four-momentum operator is defined by
\begin{eqnarray}
\hat P^{\mu}_{\mathrm {int}}&=&\int \mathrm dx^{\mu}\hat \mathcal {L}_{\mathrm{int}}(x)\nonumber\\
&=&\int \mathrm d^4 x\delta(x\cdot x-\tau^2)\theta(x_0)x^{\mu}\hat \mathcal {L}_{\mathrm{int}}(x)
\end{eqnarray}
and satisfies the point-form equations.  Further, the total
four-momentum operator, $\hat P^{\mu}:=\hat P^{\mu}_{\mathrm{free}}+\hat P^{\mu}_{\mathrm {int}}$ also
satisfies the point-form equations.

As an example consider the gluon self-coupling interaction:
\begin{eqnarray}
\hat F^{\mu\nu}_a (x)&=&\frac{\partial \hat G^{\nu}_a}{\partial
x_{\mu}}-\frac{\partial \hat G^{\mu}_a}{\partial x_{\nu}}
-\alpha c_{abc}\hat G^{\mu}_b (x) \hat G^{\nu}_c (x),\\
\hat T^{\mu\nu}(x)&=&\hat F^{\alpha\beta}_a(x)\eta_{\beta\beta'}
[\eta^\mu_{\alpha'}
\eta^\nu_{\alpha}+
\eta^\nu_{\alpha'}\eta^\mu_{\alpha}-
\frac{1}{2}\eta^{\mu\nu}\eta_{\alpha\alpha^{'}}]\hat F^{\alpha'\beta'},\\
\hat P^\mu_{\mathrm{gluon}}&=&\int_{\mathrm{hyper}} \mathrm dx_\nu \hat T^{\mu\nu};
\end{eqnarray}
here $\hat F^{\mu\nu}_a (x), \hat T^{\mu\nu}(x)$ are the field tensor and
stress energy tensor respectively, and the four-momentum operator
is integrated over the forward hyperboloid.  Also,
 $c_{abc}$ are the color structure constants and $\alpha$ is the strong
 bare coupling constant.

 Given the total four-momentum operator as a sum of free and interacting
 operators, the goal is to find the vacuum, one-particle and scattering states.
 The vacuum problem is to find a state $|\Omega\rangle$ such that
it carries a one-dimensional representation of the Poincar\'e
group and is invariant under internal symmetries:
\begin{eqnarray}
\hat P^{\mu}|\Omega\rangle&=&0\\
\hat U_{\Lambda}|\Omega\rangle&=&|\Omega\rangle\\
\hat U_c|\Omega\rangle&=&|\Omega\rangle,
\end{eqnarray}
where $c$ is an element of the internal symmetry group.

Several points can be made about the vacuum: First, it is not
possible to add a multiple of the identity operator to $\hat
P^{\mu}$ without violating  point-form equations.  This means the
vacuum state cannot be shifted to give zero energy and momentum.
Second, it suffices to calculate $\hat P^0|\Omega\rangle=0$, for
invariance under Lorentz transformations implies $\hat
{\vec{P}}|\Omega\rangle=0$.  Finally, writing
$|\Omega\rangle=e^{\hat S}|0\rangle$,  generates a set of vacuum
equations. For  simple one-dimensional models these vacuum
equations can be used to transform away the gluon self-coupling
terms;  it is not known whether the same procedure also works for
the full space-time vacuum equations~\cite{Murphy:2009}.

One of the main advantages of the point form is its explicit
Lorentz structure.  Candidates for the vacuum state are greatly
restricted by the condition that they be Lorentz scalars.
Similarly one-particle states must transform properly under
Lorentz transformations.  There are simple models where the
four-momentum eigenstates can be solved exactly because of this
transparent Lorentz structure.

 To conclude this section we describe the point-form interaction picture
 and  its accompanying covariant perturbation theory.   Starting with the
 relativistic Schr\"{o}dinger equation,  write:
\begin{eqnarray}
i\partial x_\mu \Psi(x)&=&(\hat P^\mu_{\mathrm{free}} +\hat P^\mu_\mathrm{int})\Psi(x);\\
\hat P^\mu_\mathrm{int}(x):&=&e^{i\hat P_{\mathrm{free}}\cdot x} \hat P^\mu_\mathrm{int} e^{-i\hat P_{\mathrm{free}}\cdot x};\\
\Psi(x)&=&\hat U(x,x_0)\Psi(x_0),\\
i\frac{\partial \hat U(x,x_0)}{\partial x_\mu}&=&\hat P^\mu_\mathrm{int}(x) U(x,x_0),\\
\hat U(x,x_0)&=&\hat I-i\int_{C(x,x_0)}  dy_\mu \hat P^\mu_\mathrm{int}(y)\hat  U(y,x_0),\\
\hat U(x,x_0)&=&\mathcal{P}e^{-i\int_{C} dy_\mu \hat P^\mu_\mathrm{int}(y)},
\end{eqnarray}
which is the starting point for covariant perturbation theory ($C$
is a contour in space-time).   Note that the four-momentum is not
conserved for intermediate states;  nor is the four-velocity
conserved for intermediate states.  This is to be contrasted with
point-form relativistic quantum mechanics, where the four-velocity
is conserved.

\section{Concluding Remarks}
\label{sec:conclusion}
Though point-form relativistic quantum theory is the  least
explored of the three common forms of relativistic dynamics, it
has several properties that make it well suited for applications
to hadronic physics.  The most important properties are the
kinematic nature of Lorentz transformations and the fact that
those operators that have interactions (the energy and momentum
operators) commute with one another.  The kinematic nature of
Lorentz transformations is particularly important for calculating
hadronic form factors.  Using a coupled channel approach we have
constructed hadronic current matrix elements that satisfy Lorentz
covariance and current conservation.  However, in using a
Bakamjian-Thomas approach, problems involving cluster separability
arise; surprisingly, by taking appropriate limits the point form
inspired form factors turn out to be closely related to front form
inspired form factors.  Further the coupled channel approach has
been used to provide more realistic models of hadronic decay
processes.

One of the goals of point-form  quantum field theory is to analyze
point-form QCD.  To that end we have constructed massless gluon
fields from irreps of the Poincar\'e group and begun investigating
model solutions for the gluon self-interactions.  In both finite
and infinite degree of freedom systems the goal is to exploit some
of the unique features of the point form.
%

\begin{acknowledgements}
E.P.B. acknowledges the support of the \lq\lq Fonds zur
F\"orderung der wissenschaftlichen Forschung in \"Osterreich" (FWF
DK W1203-N16).
\end{acknowledgements}



\end{document}